\documentclass[%
reprint,
groupedaddress,
showpacs,preprintnumbers,
 amsmath,amssymb,
 aps,
pra,
]{revtex4-1}

\usepackage{graphicx}
\usepackage{dcolumn}
\usepackage{bm}

\usepackage[dvips]{hyperref} 
\hypersetup{
			breaklinks=true,
			pdftitle={Optimal coupling of cold atoms in a fiber by a blue detuned hollow beam funnel},
			pdfauthor={Jerome Poulin},
			pdfkeywords={},
			pdfhighlight=/O,
			pdfstartview=FitH, 
			bookmarksopen=true, 
			bookmarksopenlevel=3, 
			pdfpagemode=UseNone,	
			pdfstartpage=1,
			plainpages=false,
			colorlinks = true,
			linkcolor = blue,
			anchorcolor = blue,
			citecolor = blue,
			filecolor = red,
			urlcolor = blue
		}

\newcommand{\fig}[3]{
\begin{figure}
\includegraphics{#1}
\caption{\label{#2}#3}
\end{figure}
}

\newcommand{\widefig}[3]{
\begin{figure*}
\includegraphics{#1}
\caption{\label{#2}#3}
\end{figure*}
}

\begin{document}


\title{Optimized coupling of cold atoms into a fiber using a blue-detuned hollow-beam funnel}


\author{Jerome Poulin}
\email[]{jerome.poulin@polymtl.ca}
\altaffiliation{Department of Engineering Physics, Ecole Polytechnique de Montreal, Montreal, H3C 3A7, Canada}

\author{Philip S. Light}
\affiliation{Frequency Standards and Metrology Group, School of Physics, University of Western Australia, WA 6009, Perth, Australia}

\author{Raman Kashyap}
\affiliation{Department of Engineering Physics, Ecole Polytechnique de Montreal, Montreal, H3C 3A7, Canada}

\author{Andre N. Luiten}
\affiliation{Frequency Standards and Metrology, School of Physics, University of Western Australia, WA 6009, Perth, Australia}


\date{\today}

\begin{abstract}
We theoretically investigate the process of coupling cold atoms into the core of a hollow-core photonic-crystal optical fiber using a blue-detuned Laguerre-Gaussian beam.  In contrast to the use of a red-detuned Gaussian beam to couple the atoms, the blue-detuned hollow-beam can confine cold atoms to the darkest regions of the beam thereby minimizing shifts in the internal states and making the guide highly robust to heating effects. This single optical beam is used as both a funnel and guide to maximize the number of atoms into the fiber. In the proposed experiment, Rb atoms are loaded into a magneto-optical trap (MOT) above a vertically-oriented optical fiber. We observe a gravito-optical trapping effect for atoms with high orbital momentum around the trap axis, which prevents atoms from coupling to the fiber: these atoms lack the kinetic energy to escape the potential and are thus trapped in the laser funnel indefinitely. We find that by reducing the dipolar force to the point at which the trapping effect just vanishes, it is possible to optimize the coupling of atoms into the fiber. Our simulations predict that by using a low-power (2.5\,mW) and far-detuned (300\,GHz) Laguerre-Gaussian beam with a 20-$\mu$m radius core hollow-fiber it is possible to couple 11\% of the atoms from a MOT 9\,mm away from the fiber. When MOT is positioned further away, coupling efficiencies over 50\% can be achieved with larger core fibers.
\end{abstract}

\pacs{37.10.Gh}

\maketitle

 	\section{\label{sec:intro}Introduction}

The field of atom optics has produced matter-wave equivalents for almost every conventional optical component \cite{Cook1982, Keith1991, Lawall1994}. However, two key elements are still missing: (i) a flexible atomic waveguide that acts as the equivalent of an optical fiber, together with (ii) the tools necessary for coupling free-space atoms into these flexible waveguides. Efficient guiding in optical fibers would allow for delivery of atoms over long distances and arbitrary paths that might enable new applications in, e.g., atom interferometry \cite{Godun2001,Andersson2002}, gravity sensors \cite{Harris1995}, ultra-precise atom implantation \cite{Ito1997,Meschede2003}, and atomic lithography \cite{Williams2006}.

In this paper, we develop a solution for both elements with a single optical beam. We consider the particular situation in which we use a repulsive potential to guide the cold atoms. This approach is interesting because the atoms can be guided in a minimally perturbating location with minimal unwanted heating. Unfortunately, as we will show, this approach has potentially much lower coupling efficiency than the opposite approach using an attractive potential. However, we will identify the conditions in which this repulsive guiding can provide the same order of efficiency as the attractive potential. We also show a rather unexpected conclusion that there is an optimal potential strength, under a given set of initial conditions, to maximize the atomic coupling. In other words, the deepest possible potential does not provide the best coupling of free-space atoms into the optical fiber.

Following the first theoretical proposals \cite{OlShanii1993,Marksteiner1994}, room-temperature atom guidance in glass capillaries was quickly demonstrated using red-detuned Gaussian beams \cite{Renn1995} as well as blue-detuned evanescent waves \cite{Ito1996, Renn1996}. In contrast to guiding with red-detuned beams, which draw atoms into the high intensity areas of the beam, blue-detuned guides present the inherent advantage of minimizing the heating of atoms due to spontaneous emission and reduce perturbation of the energy levels by the light potential (Stark shifting) \cite{Marksteiner1994}. Shortly after these first demonstrations, evanescent guiding of laser-cooled atoms using beams of He \cite{Dall1999} and Rb \cite{Muller2000} was achieved. Further progress was limited because optical guidance in capillaries is both lossy \cite{Renn1997} and highly multimode \cite{Dall1999} which leads to high atomic losses and short guidance lengths ($\sim$cm). One of these models \cite{Yin1998} also proposed the use the diffracting field from the cladding to couple the atoms, although the low power in the evanescent field required a high-power near-detuned beam with its associated spontaneous emission losses. 

Development of hollow-core photonic-crystal fibers (HC-PCF) \cite{Cregan1999, Roberts2005} paved the way for efficient guidance of atoms because of their low loss and single mode optical guiding. The first demonstration of room-temperature atoms guiding through a HC-PCF reported a guiding efficiency of up to 70\% through a 6-cm-long fiber \cite{Takekoshi2007}. Following soon after, two experiments demonstrated the loading of cold atoms into HC-PCF using red-detuned Gaussian beams \cite{Christensen2008, Bajcsy2009}. Very recently, cold Rb atoms were guided through a 8.8-cm-long HC-PCF by a 2.3\,Watt far-detuned Gaussian beam and a peak flux of $10^5$\,atom/s was observed \cite{Vorrath2010}. 

Two approaches were recently developed to improve the coupling efficiency of cold atoms into small-core HC-PCF. Bajcsy \emph{et al.} used a quadrupole magnetic funnel to guide the MOT from its launch distance to within reach of the Gaussian beam funnel \cite{Bajcsy2009}. Soon after, they improved the atom coupling efficiency by a factor of 6 using a collimated hollow-beam and a repulsive sheet beam. The system effectively acts as an ``optical elevator" that lowers the MOT within 1\,mm of the fiber input \cite{Bajcsy2011}. When the sheet beam is turned off, gravity and a red-detuned Gaussian beam funnels the atoms into the core. 

For blue-detuned atomic guidance, HC-PCF enables the use of a hollow Laguerre-Gaussian beam itself to guide the atoms instead of just the evanescent-field part that leaks into the core. This allows for a deeper trapping potential at much lower optical power, while keeping the atoms away from influence of the fiber walls \cite{Marksteiner1994,Sague2007}.  Collimated Laguerre-Gaussian (LG) beams have also been used to guide cold atoms from a MOT and it was shown that high $\ell$-order beams proved to be the most efficient guides as they minimize the potential energy given to atoms at the loading stage \cite{Mestre2010}. However, to our knowledge, no HC-PCF have been demonstrated to guide high $\ell$-order LG beams. On the other hand, low-loss guidance of a LG$_{01}$-like mode in a 19-m-long HC-PCF was recently demonstrated in \cite{Euser2008}. Models of atom guiding in collimated blue-detuned hollow-beams in vacuum have also been developed and validated with experimental results \cite{Xu1999, Xu2000, Mestre2010}, although to our knowledge no one has yet considered the cold atoms behavior in the diffracted field from the core of the fiber.

In this paper, we present the solution to the best parameters for optimal atomic coupling. In Sec.~\ref{sec:theory}, the physical equations and 3D Monte Carlo numerical model of our proposed experiment are presented. Sec.~\ref{sec:dynamics} deals with the dynamics of the atoms in the funnel and the gravito-optical bottle trapping effect. The system's atom coupling efficiency against experimental parameters is studied in Sec.~\ref{sec:system}.

\section{\label{sec:theory} Atom coupling model}

We propose, in Fig.~\ref{fig:setup}, an experimental arrangement in which a far-detuned low-order Laguerre-Gaussian beam is coupled and guided in a hollow-core photonic-crystal optical fiber \cite{Cregan1999}. %
\fig{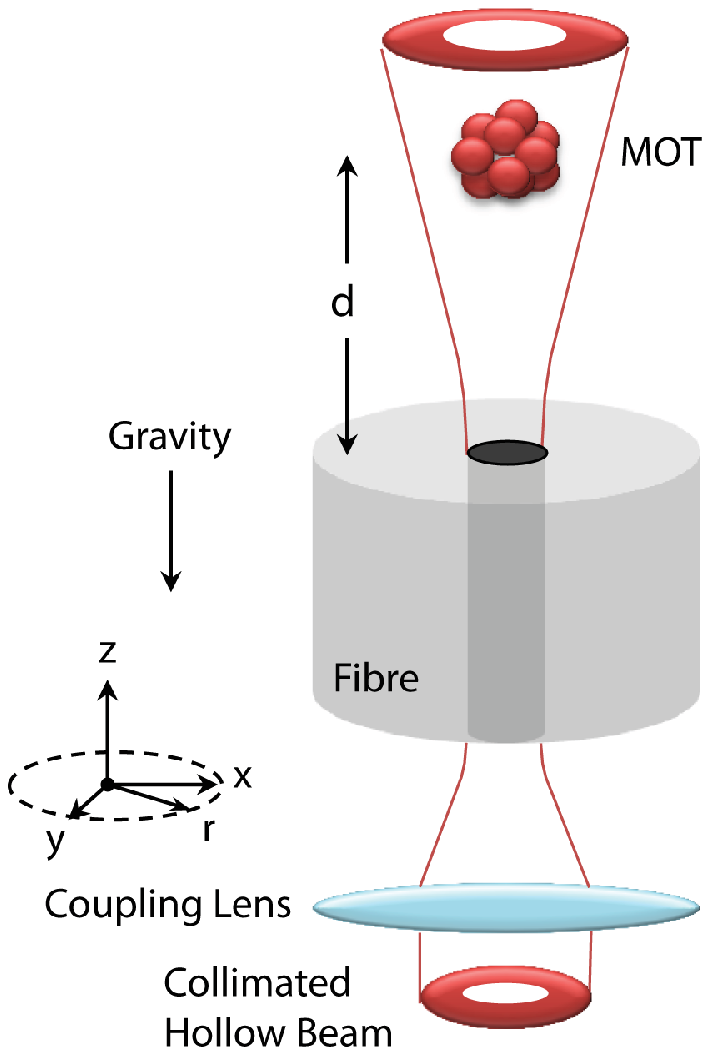}{fig:setup}{(Color online) Experimental setup for coupling atoms from a MOT into a hollow-core photonic-crystal fiber. The fiber is mounted vertically in the vacuum chamber. A single blue-detuned collimated Laguerre-Gaussian beam is coupled into the fiber core at the bottom of the fiber. At the top end, the beam diffracts into free space acting as a funnel. When the MOT is released, from distance \emph{d} of the fiber output, the combined action of gravity and the hollow-beam funnels the atoms into the core and guides them down the fiber. The schematic is not drawn to scale.}%
The fiber is aligned vertically, so that its axis is parallel to gravity. A blue-detuned hollow-beam diffracting from the core acts as a funnel to both couple and guide atoms released from a MOT.

In this paper, the term ``coupling" is used to describe the process of optically funnelling the atoms into the fiber core. The term ``guiding" is exclusively used to describe the optical control of the motion of atoms when they are inside the fiber core.

\subsection{\label{sec:derive_forces} Forces in the optical funnel}

The rotational symmetry allows the problem to be analysed using cylindrical coordinates: \emph{r} being the distance to the fiber core central axis and \emph{z} the axial distance from the fiber. The extremity of the fiber facing the MOT defines the axial origin ($z=0$). We will use this form to describe the intensity of a single-ringed Laguerre-Gaussian mode (radial order $\rho=0$, azimuthal order $\ell\neq0$) \cite{Friedman2002}:
\begin{equation}
I\left( {r,z} \right) = \frac{{2P}}{{\pi w{{(z)}^2}}}\frac{1}{{\left| \ell \right|!}}{\left( {\frac{{2{r^2}}}{{w{{(z)}^2}}}} \right)^{|\ell|}}exp\left( {\frac{{ - 2{r^2}}}{{w{{(z)}^2}}}} \right)
\label{eq:intensity}
\end{equation}
where \emph{P} is the optical power and $w(z) = w_0 \sqrt {1 + ({z \mathord{\left/
 {\vphantom {z {z_R )^2 }}} \right.
 \kern-\nulldelimiterspace} {z_R )^2 }}} $ is the radius of the lowest order beam propagating in free space. The beam waist radius, inside the fiber, is $w_0$ and $z_R = \pi \omega_0^2 / \lambda$ is the Rayleigh range. When the laser frequency is far-detuned ($>$1000 $\Gamma$, $>$40\,GHz for Rb atoms), only the ``off-resonant" dipolar force is effective and the trapping potential is expressed in the two-level approximation \cite{Balykin1995}: %
\begin{equation}
U(r,z) \simeq \frac{{\hbar {\Gamma ^2}}}{{8\delta }}\frac{I}{{{I_s}}}
\label{eq:farpotential}
\end{equation}
where the laser detuning $\delta=\omega_{laser}-\omega_0$, $\Gamma$ is the natural linewidth of the cooling transition, and $I_s$ is the saturation intensity parameter. In this paper, we will specify the energies of atoms in terms of an equivalent thermal energy for ease of comparison between energy scales. The dipolar force is equal to the negative gradient of the potential $U(r, z)$.  In the interest of readability, the vector force will be separated into its axial ($F_{z}$) and radial ($F_{r}$) components. Explicit mention of the variable's dependence in \emph{r,z} will also be omitted. Deriving the trapping potential [Eq.~(\ref{eq:farpotential})] in the context of the LG intensity [Eq.~(\ref{eq:intensity})], in \emph{r} and \emph{z}, we obtain:
\begin{align}
F_r& = \frac{-\hbar \Gamma ^2}{8\delta I_s } \frac{{{2^{\ell + 2}}P}}{{\pi \ell!}}\frac{{{r^{2\ell}}}}{{{w^{2\left( {\ell + 1} \right)}}}}\left[ {\frac{\ell}{r} - \frac{{2r}}{{{w^2}}}} \right]
 exp\left( {\frac{{ - 2{r^2}}}{{{w^2}}}} \right)		
 \label{eq:tforce}
  \\[6pt]
F_z& = \frac{-\hbar \Gamma ^2}{8\delta I_s } \frac{{ - {2^{\ell + 2}}P}}{{\pi \ell!}}\frac{{{r^{2\ell}}}}{{{w^{2\ell + 3}}}}\left[ {1 + \ell - \frac{{2{r^2}}}{{{w^2}}}} \right]\nonumber
\\
&\qquad\qquad\qquad \cdot exp\left( {\frac{{ - 2{r^2}}}{{{w^2}}}} \right)\frac{{{w_0} \cdot z}}{{{z_R}^2\sqrt {1 + {{\left( {z/{z_R}} \right)}^2}} }}
\label{eq:axforce}
\end{align}
As an example, when comparing the axial and transverse components of the dipolar force at the same spatial coordinates in an LG$_{01}$ beam with a $1/e^2$ waist diameter of 40\,$\mu$m (diffraction half-angle of 1.6$^\circ$), we find that $F_z$ is at least 3 orders of magnitude weaker than $F_r$. Nonetheless, $F_z$ can be many times stronger than gravity, even for a realistic trap potential depth of 0.5\,mK.  This results in $F_{z}$ having a surprisingly strong influence on the dynamics of the atoms in the trap.  It can be seen from Eq.~(\ref{eq:farpotential}) that for large detunings, the potential is merely linearly dependent upon the ratio of the optical power over the detuning. From now on, we will refer to this parameter as the light force parameter $\kappa = P/\delta$.

The radial force confines atoms within the funnel as gravity pushes them downwards toward the fiber core. Fig.~\ref{fig:tforce} shows the radial acceleration on  $^{85}$Rb resulting from a dipolar radial force under typical conditions at three axial distances from the end of the fiber. The force has the form of a trapping potential, which has its maximum at the output of the fiber and decreases with distance proportional to the intensity. An important consequence of the intensity distribution of a blue-detuned hollow-beam is its limited confinement radius. Outside the maximum intensity radius, the decreasing gradient produces a force that repels atoms away from the axis of the beam. %
\fig{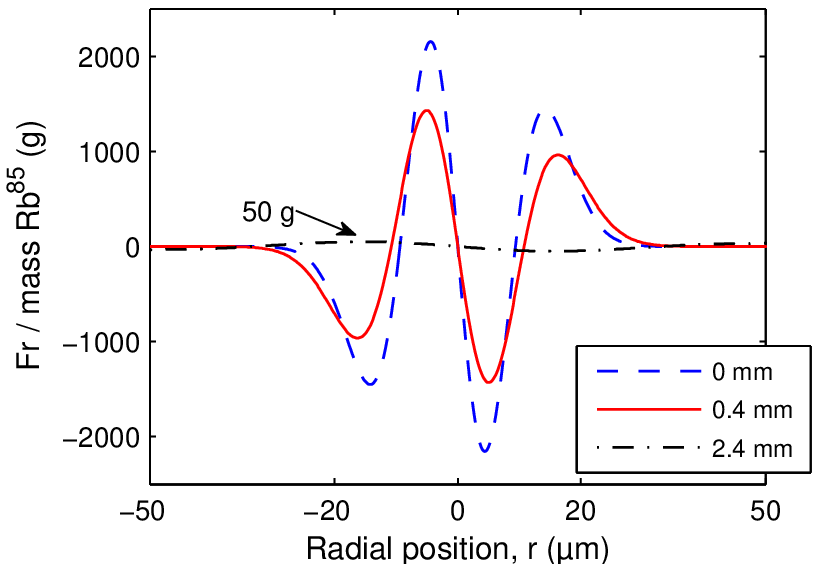}{fig:tforce}{(Color online) Radial component of the blue-detuned dipolar force [see Eq.~(\ref{eq:tforce})] at three distances from the fiber end: 0\,mm (blue dashed), 0.4\,mm (red solid) and 2.4\,mm (black dash-dotted). The force is represented as $^{85}$Rb acceleration in units of standard gravity. Positive values means the force vector is oriented towards increasing values of \emph{r} (right), while negative values indicate the force vector points to decreasing values of \emph{r} (left). This calculation used a 4\,mW $LG_{01}$ beam, blue-detuned to 300\,GHz from the $^{85}Rb$ $D_2$ cooling transition ($5^2S_{1/2} (F=3) \rightarrow 5^2P_{3/2} (F=4)$) around 780.24\,nm and diffracting from a 40-$\mu$m-diameter fiber core. The transition natural linewidth is $\Gamma/ 2\pi = 5.89$\,MHz and saturation intensity is $I_{sat} = 1.63$\,mW/cm$^2$ \cite{Steck2009}. } %

The axial force, $F_{z}$, is a manifestation of the gradient of intensity produced by a diverging beam. Therefore, it is null for collimated beams, e.g. inside the fiber. The axial force is always directed away from the end of the fiber for a blue-detuned beam regardless of the optical beam's propagation direction. This force is seen to rise sharply when the atom comes within 1.5\,$z_R$ of the end of the fiber. Under typical coupling conditions, the axial force will be many times the gravitational force when close to the radius of maximum intensity. As an example, Fig.~\ref{fig:axforce} shows the axial acceleration on  a $^{85}$Rb atom  for three axial distances, \emph{z}: 0.02, 0.4 and 2.4\,mm. The axial acceleration reaches a maximum at 0.4\,mm. At distances of 0.02 and 2.4\,mm, the axial force maximum is equal to the gravity.%
\fig{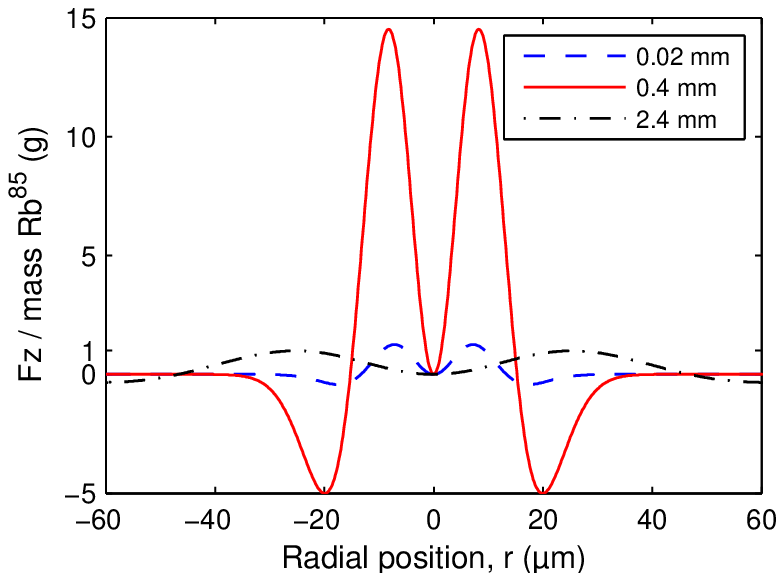}{fig:axforce}{(Color online) Axial component of the blue-detuned dipolar force [see Eq.~(\ref{eq:axforce})] at three distances from the fiber end: 0.02mm (blue dotted), 0.4 mm (red solid) and 2.4 mm (black dashed). The force is represented as $^{85}Rb$ acceleration in units of standard gravity. Positive values means repulsive while negative is an attractive force. The same physical parameters have been used as in Fig.~\ref{fig:tforce}.}

\subsection{Monte Carlo simulation model}

A key objective of this research was to identify the best coupling efficiency using blue-detuned light with realistic representations of the initial state.  The coupling efficiency is defined as the ratio of the number of atoms that enter the fiber core over the total number of atoms loaded in the MOT. The Monte Carlo method was used to determine an initial state for each atom in the MOT. The atoms are treated as independent point particles in the MOT and we used a Maxwell-Boltzmann distribution for their velocities \cite{Metcalf1999}. We selected a temperature of 25 $\mu$K, which can be achieved with careful tuning of a standard MOT \cite{Wallace1994}. We modelled the density variation as Gaussian ($\rho(r)=\rho_0 exp(-r^2/\sigma^2)$) with the 1/e radius of the trap of 70 $\mu$m, which can be experimentally achieved with careful tuning of the optical and magnetic forces to optimize the trap spring constant \cite{Wallace1994,Petrich1994}.

We numerically simulate the trajectory of each atom in the MOT from the time-of-release. The Velocity-Verlet algorithm was selected because of the algorithm's property to conserve the energy of particles modelled in a conservative system \cite{Garcia2000}.  At every timestep $\Delta t$, the position and velocity of each atom is calculated as follow:
\begin{align}
r(t+\Delta t)& =r(t)+v(t)\Delta t+ \textstyle{\frac{1}{2}} a(t)\Delta {t^2} \label{eq:vvr} \\
v(t+\Delta t)& =v(t)+\textstyle{\frac{1}{2}} \left[a(t)+a(t+\Delta t)\right]\rm{\Delta }t  \label{eq:vvv}
\end{align}
The accelerations $a(t)$ and $a(t+\Delta t)$ are both simply the sum of the accelerations by gravity and the dipolar force Eqs.~(\ref{eq:tforce}),(\ref{eq:axforce}) at their corresponding positions in time $r(t)$ and $r(t+\Delta t)$. Errors in the simulation can be readily estimated by observing changes in the particle energy during the simulation.

	\section{\label{sec:dynamics} Atom coupling dynamics}

In this section, we undertake a detailed examination of the atom trajectories to understand the behaviour of the atoms. We distinguish the outcome into three classes: (i) atoms coupled into the fiber (``coupled"), (ii) those captured in the funnel but which do not enter the fiber (``trapped"), (iii) and those which immediately escape the funnel when released from the MOT (``escaped"). In Sec.~\ref{sec:motion}, we will analyze atomic motion and then, adding the particular details of a MOT released into an optical blue-detuned funnel in Sec.~\ref{sec:atomdynamics}.

\subsection{\label{sec:motion}Motion of atoms in an optical dipolar potential}

By considering atoms inside the fiber, whether guided in a blue-detuned collimated Laguerre-Gaussian beam or red-detuned Gaussian beam, their motion will be that of a particle trapped in a central, attractive and circular potential similar to the case of a test mass moving in the gravitational field of a point mass. More generally, for a central power-law potential $V = r^x$, only potentials in $x = -1$ (gravity) or $x = 2$ (harmonic) can sustain stable closed elliptical orbitals \cite{Danby1988}. The optical confining potential follows a sigmoidal-shaped function of $r^2 e^{-r^2}$. Therefore, we expect that atoms guided/loaded in the optical dipolar potential will not describe stable orbitals, which we do find, as shown in Fig. \ref{fig:path}. This figure displays a characteristic orbital for an atom falling into the funnel.

\fig{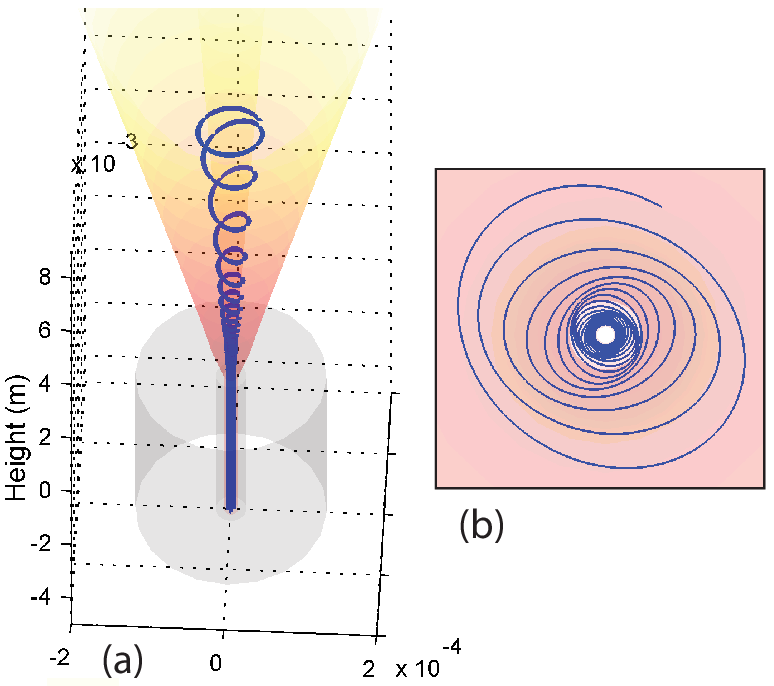}{fig:path}{(Color online) (a) A characteristic orbital trajectory of an atom captured by the confining potential and coupled into the fiber. (b) Top-view of the elliptical orbital trajectory as the atom is guided toward and inside the fiber. By conservation of  angular momentum, the atom accelerates transversally as its elliptical movement is compressed into an ever smaller orbital.}
If we consider an atom falling into a diffracting potential, the atom will spiral into the light funnel, analogous to a ball moving on a hyperbolic funnel surface. As the diameter of the potential reduces towards the waist of the beam, the orbital velocity of the moving particle is seen to increase as is expected from conservation of the angular momentum. Unlike the ideal circular motion in the classic hyperbolic funnel, we see the orbitals describe open precessing ellipses as seen in Fig. \ref{fig:path}. Similar trajectories are also observed in collimated beam simulations. Depending on the initial conditions, simulated orbitals range from almost circular to extreme ellipses (close to a purely radial motion).

In the optical funnel, the axial motion is linked to the transverse velocity in a complex fashion.   In the absence of axial forces in the funnel, i.e. where the diffraction is absent or inside the fiber, all atoms accelerate downwards with gravity independently of their orbital motion. However, in locations where diffraction is significant, the tangential velocity assumes much greater importance.  The orbital radius of a high tangential velocity atom needs to be larger in order to be balanced by the stronger radial forces available at that location.  In this case the atom will penetrate further into the high intensity areas of the beam and thus, the effective vertical acceleration delivered by the hollow-beam will be stronger. By this means, the optical funnel can couple tangential and axial motions of the atom and yield the complex orbits that are seen on Fig.~\ref{fig:path} as well as the axial velocities illustrated in Fig.~\ref{fig:velocities}.

The path integral of the axial force, along the atom trajectory in the optical funnel, reveals the potential barrier that needs to be overcome in order to reach the fiber.  Consequently, all atoms funnelled toward the fiber experience some deceleration that is dependent on their path trajectory. It is this complex interaction of their initial state energies (transverse, axial and gravity potential) with the varying forces of the optical dipolar funnel which determines whether they couple into the core. %
Fig. \ref{fig:velocities} presents the axial velocities of each atom as a function of its distance from the fiber, in a Monte Carlo atom coupling simulation. The atoms trajectories have been labelled on this figure according to the three possible outcomes: escaped from the funnel (green), guided into the fiber (blue), or trapped in a dynamic balance between gravity and the axial dipolar force (red). Once the guided atoms have gone through the repulsive axial potential barrier, they all start accelerating again inside the fiber by the action of gravity. The ``trapped" atoms are pushed upwards and are recycled into the same near-triangular path of velocity-distance, being effectively trapped by the conservative potential. %
\fig{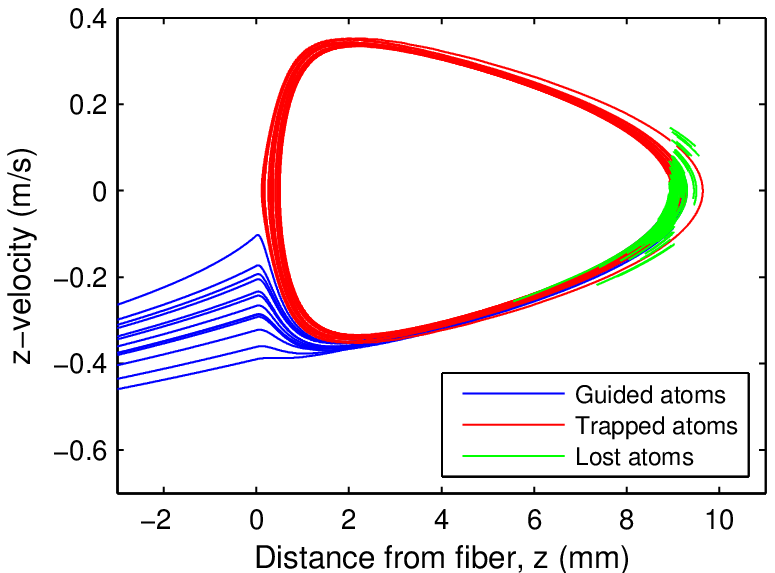}{fig:velocities}{(Color online) The axial velocity of each atom in the simulation is represented against its distance from the fiber. All atoms are released from 9\,mm and exposed to a 4\,mW/300\,GHz blue-detuned LG$_{01}$ beam with a 20-$\mu$m waist radius within the fiber. Negative distances represent positions within the fiber core from the coupling input. Each velocity path is color-coded by the final outcome of its atom: guided (blue), trapped (red), and escaped (green).}

\subsection{\label{sec:atomdynamics}Atom dynamics in a microscopic optical funnel}

\widefig{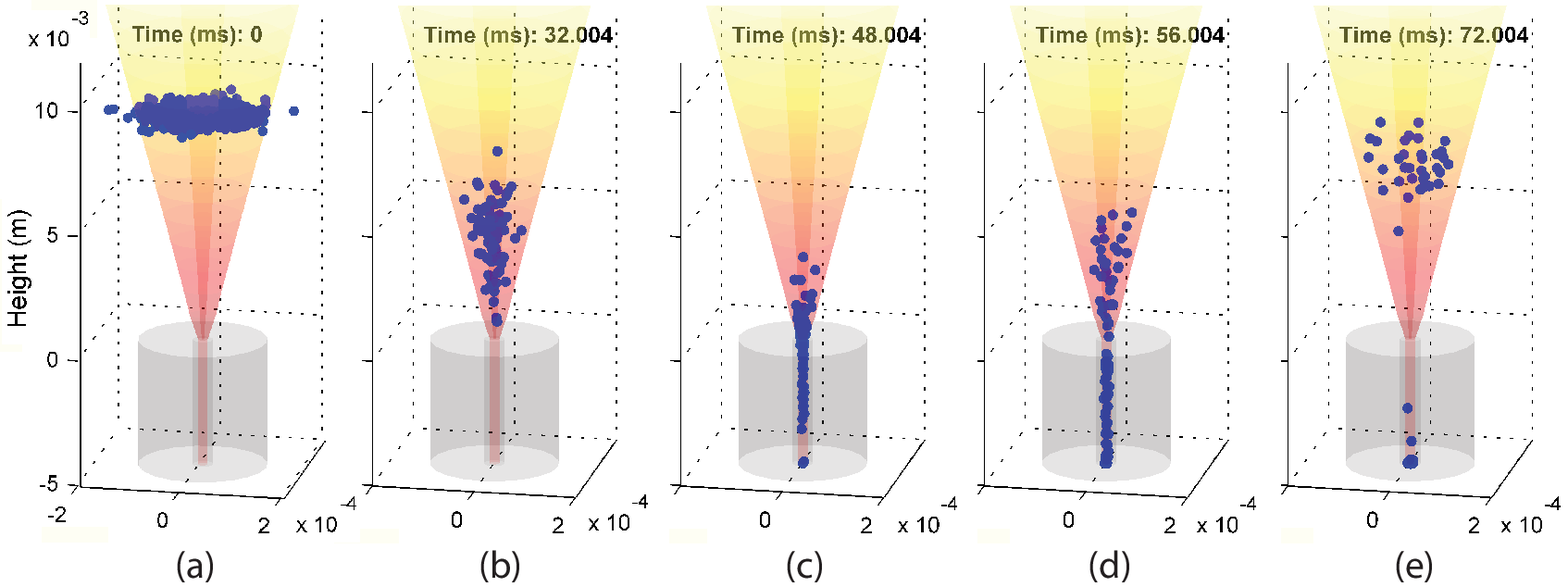}{fig:movie}{(Color online) The dynamics of atomic coupling are represented with 5 snapshots taken during the simulation. The vertical axis is compressed by a factor of 15. The limits of confinement are represented by the outer light cone. The inside cone is delimited by the dark radius (peak intensity/e$^2$). (a) Atoms at release time from the cooling lasers, 9\,mm away from the fiber. Their density distribution is a 3D Gaussian but the compressed vertical scale makes it appear like a disk. (b) 16\% are loaded in the hollow-beam funnel. Their phase space is compressed as they approach the fiber. (c) Almost all atoms approach the fiber within 0.5\,mm, with some of them already coupling into the core. (d) Atoms are separated into two energy-selective populations: the coldest 8\% are guided into the fiber core, the other 8\% are pushed away by the axial component of the dipolar force. (e) Atoms relax up until gravity forces them down again toward the fiber. This cloud of atoms is trapped in the gravito-optical bottle trap as long as their total energy is conserved in the system.}

Fig.~\ref{fig:movie} displays five sequential snapshots of a 3D Monte Carlo simulation to fully illustrate the dynamics described. The figure is vertically compressed by a factor of 15 because the \emph{z} axis spans over 17\,mm while \emph{x} and \emph{y} axis are only 0.4\,mm. In Fig.~\ref{fig:movie}(a), the initial spatial distribution of atoms is a 3D Gaussian function (spherical); although they initially appear to be in a disk distribution due to the compressed \emph{z} scale aforementioned. The confinement area is illustrated by the red cone colored from light yellow to red following the increasing trap potential. In our simulations, we observed that the minimal efficient coupling distance requires the radius of the maximum intensity of the beam, at MOT release location, to reach 1.8 times the 1/e-fold width of the MOT [as is the case in Fig.~\ref{fig:movie}(a)]. This effectively corresponds to releasing from the MOT $>$98\% of the atoms within the confinement limits of the funnel.

Atoms initially outside the red cone will be immediately lost. Those atoms inside the cone can escape only if they possess a transverse energy \emph{E$_{Tr}$} greater than the trap potential depth, \emph{$U_0(z)$}. The transverse energy of an atom at position \emph{x, y, z} is defined:
\begin{equation}
E_{Tr}(x,y,z) = E_{Kt}(v_x,v_y) + U(x,y,z)
\label{eq:Etr}
\end{equation}
where $E_{Kt}$ is the kinetic energy in the plane perpendicular to the fiber/beam axis and $U(x,y,z)$ is the optical trap potential at the atom position calculated from Eq~(\ref{eq:farpotential}).

After the initial loss of the higher energy atoms in the distribution, in Fig.~\ref{fig:movie}(b), the remaining cloud will no longer exhibit a Maxwell-Boltzmann distribution. We observe an areal compression of the cloud as it falls in the funnel. Although the total energy of each atom is conserved as it falls, the conversion of gravitational potential energy into kinetic energy leads to a large increase in the speed of the atoms. The diffraction of the funnel causes a conversion of the gain in axial velocity into an increase in the transverse energy. The simulation shows that the increase in the transverse energy is proportional to the decrease in the area defined by the locus of points of maximum intensity. That is, for a typical situation, with the MOT positioned 9\,mm of the fiber, we find the radius of maximum confinement decreases by 160 times from the MOT release to the entrance of the fiber. The transverse energy of each atom also increases by 160 times. Fortunately, the trapping potential increases proportionally to the square of the decreasing fiber distance, which exactly matches the dependence of the transverse energy keeping the atoms in a trap of the same relative depth (ie. max transverse energy/optical potential depth is approximately constant).

Fig.~\ref{fig:movie}(c) shows the atoms falling toward the fiber in a cone-shaped cloud, coming very close to the fiber surface, until a separation into two populations occurs [Fig.~\ref{fig:movie}(d) and (e)]. The coldest fraction of atoms couples into the core; while others are repelled from the fiber.  Continuation of the simulation shows that these atoms are executing the close-cycle trajectories shown on Fig.~\ref{fig:velocities} as the  ``trapped" atoms.

\fig{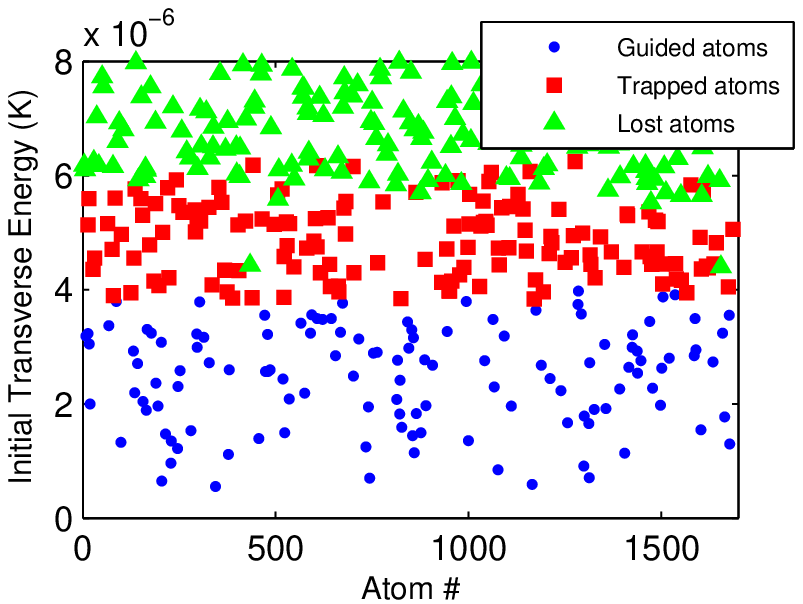}{fig:tenergy}{(Color online) Initial transverse energy of 1700 $^{85}$Rb atoms labelled in function of their simulation results: guided into the fiber (blue dots), trapped by the gravito-optical bottle effect (red squares) or escaped the hollow-beam (green triangles). The 3D Monte Carlo simulations had the following parameters: 25\,$\mu$K, 200\,$\mu$m diameter MOT launched at 9\,mm into a LG$_{01}$ beam of 40\,$\mu$m diameter, 3\,mW and 300\,GHz detuning.  Only escaped atoms with transverse energy below 8\,$\mu$K are shown.}%
The principal determinant in the final state of the atoms is the initial transverse energy of the particle. The initial transverse energy of 1700 $^{85}$Rb atoms was compared against their outcome in Fig.~\ref{fig:tenergy}. It is shown that the coldest atoms (with lower initial transverse energy) couple into the fiber core (blue circles). Atoms with higher energy immediately escape the hollow-beam at the MOT release distance (green triangles). Atoms with transverse energy intermediate to those values penetrate the confining potential more deeply (red squares). These atoms have their axial velocity reversed before they reach the fiber. These atoms end up being trapped in an energy-selective gravito-optical bottle trap. 

Similar gravito-optical traps based on blue-detuned hollow-beams have been previously reported \cite{Ovchinnikov1998,Yin1998a}, although their intention was to trap the atoms whereas we are trying to avoid this. Under our set of conditions, the axial component of the blue-detuned repulsive force, opposing gravity, is only effective for atoms with higher transverse energy. During the coupling experiment, the gravito-optical bottle trap effectively behaves as a hot-atom filter. As the potential is increased in strength a larger fraction of atoms fall into this ``trapped" class. Obviously, this bottle trap effect should be carefully managed in order to achieve optimal coupling of atoms into the fiber core.

One can see from Fig.~\ref{fig:tenergy}, that the transverse energy is not the singular determinant of the eventual classification of the atom. We find that the initial axial velocity can have a minor influence for those atoms with axial velocities at the extrema of the distribution. This causes an atom to fall into a class which was unexpected from their initial transverse energy alone. In Sec.~\ref{sec:system}, we will consider how the experimental parameters influence the atom's distribution among the three classes.

 	\section{\label{sec:system}System properties: optimizing coupling efficiencies}

In this section, we will focus on evaluating the coupling efficiency of our specific arrangement. In particular, we will discuss the dependence of the efficiency on: the light force parameter, MOT-fiber distance and fiber core diameter. We will also make a comparison of this hollow-core approach with an optimal coupling scheme based on red-detuned Gaussian beam coupling.

\subsection{\label{sec:optimal}Optimal coupling efficiency discussion}

In our simulations with $^{85}$Rb, we selected realistic MOT parameters of T=25\,$\mu$K and the trap radius $\sigma=70 \mu$m using a standard setup \cite{Wallace1994}. We have centred the MOT on the fiber axis and made use of a HC-PCF that guides a 40-$\mu$m diameter LG$_{01}$ beam. The remaining degrees of freedom are the dipolar light force parameter ($\kappa = P/\Delta$) and the distance between the fiber and the centre of the MOT.
 \fig{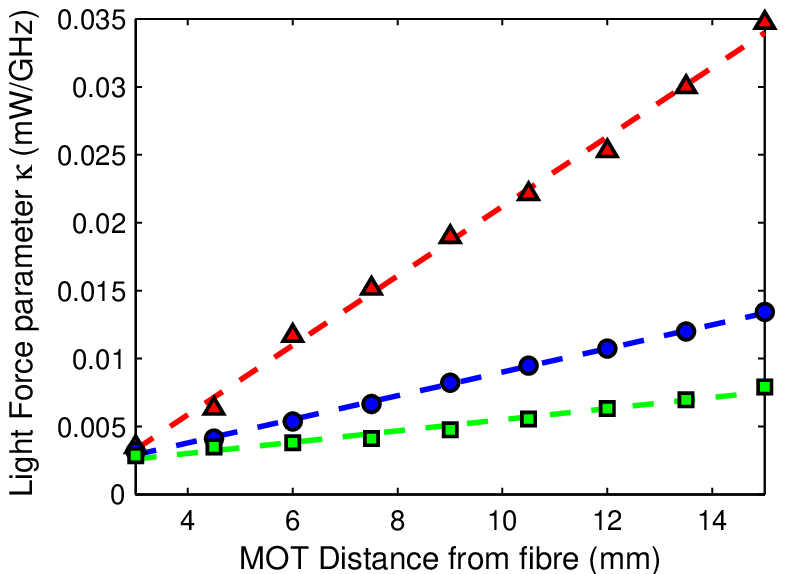}{fig:optimal}{(Color online) Atom coupling efficiency as a function of the force parameter of the blue-detuned hollow-beam and the distance the atoms in the MOT are released from the fiber. The optimal line (blue circles) shows the force relation with MOT distance to realize the best coupling efficiency. The low-force line (green squares) and high-force line (red triangles) represent the force-distance combination that would deliver a factor 2 reduction in the coupling efficiency.}
The simulation shows that there is an optimal potential for maximum coupling at a given fiber-MOT spacing.   This optimal value is shown by the blue circles line on Fig.~\ref{fig:optimal}. We also show two other lines in which the coupling fraction has fallen by a factor of two from the optimal value. The relationship between the force necessary to maintain optimal coupling and the MOT distance is linear. This arises because we need to compensate for the additional diffraction by increasing the beam power to maintain the equivalent initial conditions. The strength of the axial potential barrier that increases proportionally with \emph{$\kappa$}, is a determinant factor in the coupling efficiency. Its effect on the coupling efficiency is compensated by the linear increase of gravitational potential energy with distance. The axial barrier stays therefore at the same relative strength for a given atom and the optimal coupling efficiency is maintained.

\subsubsection{\label{sec:cpldistance}Optimal coupling as a function of the MOT-fiber distance}

\fig{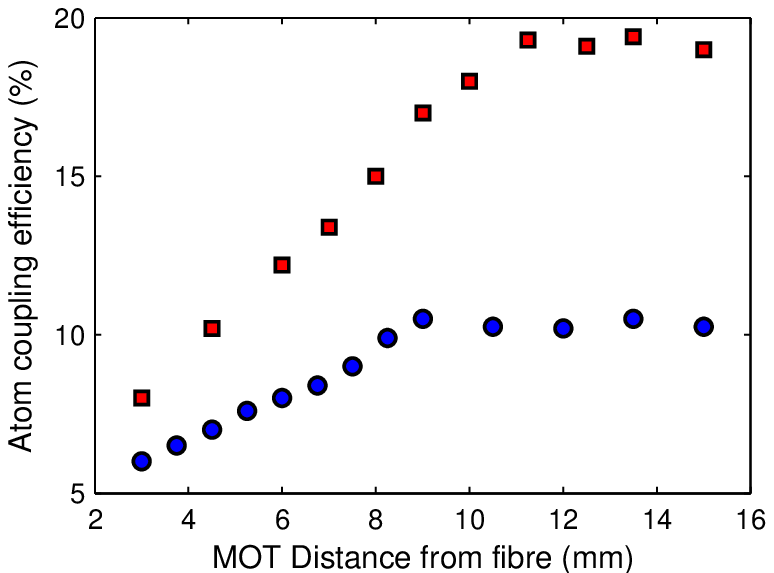}{fig:bestcoupling}{(Color Online) Monte Carlo simulation of the optimal coupling efficiency of a 25\,$\mu$K, 200\,$\mu$m diameter $^{85}$Rb MOT launched from a fixed distance into: a 40-$\mu$m diameter (blue circles) and 50-$\mu$m diameter HC-PCF (red squares) guiding a blue-detuned LG$_{01}$ beam.} 
Fig.~\ref{fig:bestcoupling} shows the maximum atom coupling efficiency obtainable as a function of the distance from the MOT for two different fiber core radii.  The analysis reveals, that beyond a minimum distance, the best coupling efficiency can be achieved. This value can be maintained for increasing distance. This distance is set when the radius of the diffracting funnel is at least 1.8 times the size of the MOT. Below this threshold the Gaussian distribution of atoms in the MOT is trimmed by the funnel confinement limits. Even if the coupling efficiency can be maintained with longer distance, it is not desirable to do so because the increased light force parameter will results in higher scattering rates in the fiber core. Therefore, it is best to work with the lowest force possible by selecting the shortest distance for optimal coupling efficiency. The minimum distance consistent with capturing the maximum fraction of the MOT is 9\,mm with a 40-$\mu$m core diameter and 11.25\,mm with 50-$\mu$m.

\subsubsection{\label{sec:cplforce}Coupling efficiency as a function of the light force parameter}

The fraction of in-coupled atoms as a function of the light force parameter in the hollow-core guide is displayed on Fig.~\ref{fig:efficiency} for our standard set of conditions. We have chosen a fixed distance of 9\,mm between the MOT and the 40-$\mu$m fiber. %
\fig{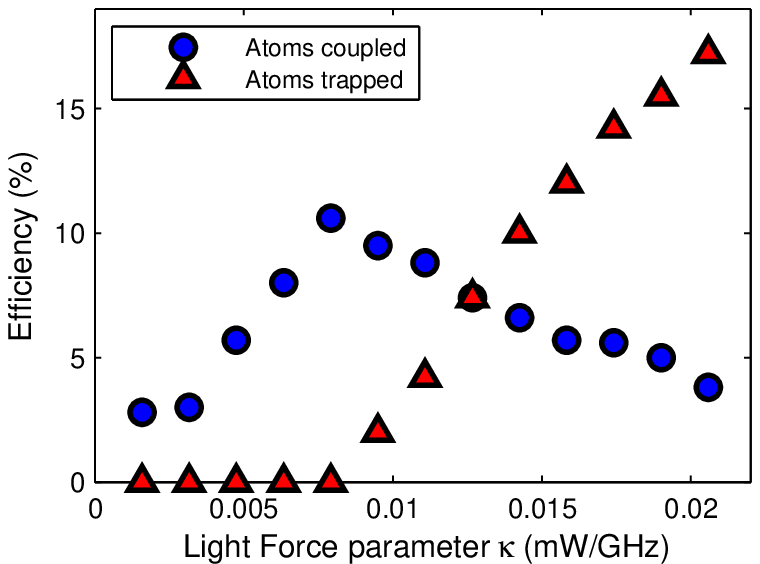}{fig:efficiency}{(Color Online) Releasing the MOT at 9\,mm from the fiber, the coupling efficiency of  $^{85}$Rb atoms is illustrated in relation to the light force parameter (blue circles). The trapping efficiency is also represented against the same light force parameter (red triangles). Best coupling efficiency is obtained with the maximum force allowed before increasing the fraction of atoms trapped in the gravito-optical bottle effect} 
The coupling efficiency increases up to a certain force, and then, decreases as the fraction of trapped atoms grows. We see that there is an optimal depth to the potential: this characteristic appears because we need to set a balance between the depth of the potential, in terms of atom capture, against the appearance of a population of trapped atoms.  The coupling-efficiency--light-force relationship shows the same optimum with different MOT release distances and different fiber core radii. Thus, for a set distance and fiber core radius, the optimal coupling efficiency is always achieved with the maximal force just before the appearance of trapped atoms.  The fraction of trapped atoms rises much faster because not only guided atoms are converted into trapped atoms but the increase in force also increases the fraction of captured atoms. Nonetheless, their higher transversal energy means they cannot be converted to coupled atoms either.

\subsubsection{\label{sec:fibcore}Scaling of coupling efficiency with fiber core diameter}

If the objective is to couple large numbers of atoms into a fiber, it is clearly an advantage to work with larger core fibers. With a larger core, there is less diffraction (diffraction angle is inversely proportional to the radius) and this results in a naturally weaker axial force. %
\fig{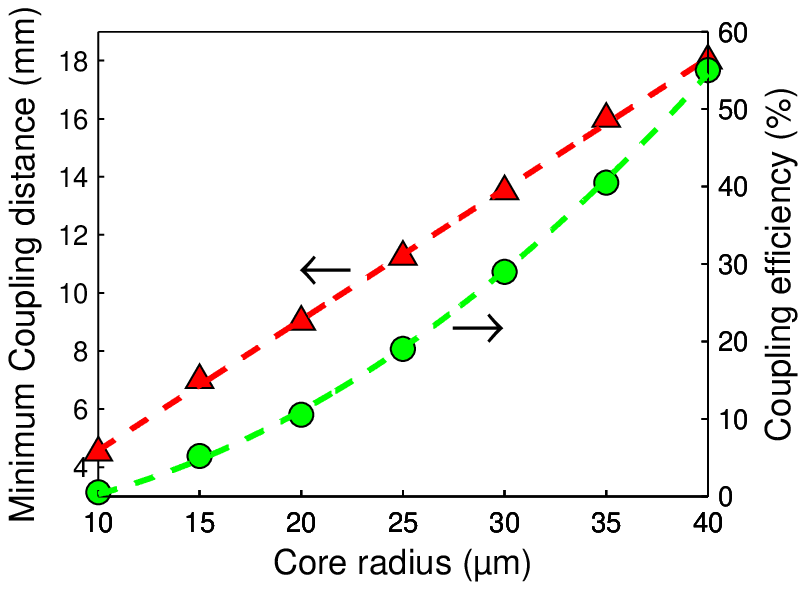}{fig:disteffcore}{(Color Online) On the left axis, minimum coupling distance to achieve the best coupling efficiency is represented as a function of the fiber core radius (red triangles).  On the right axis, the best coupling efficiency achieved with a particular fiber core radius is shown (green circles). The red dashed line is a linear fit of the distance-radius relationship and the green dashed line is a quadratic fit to the coupling efficiency.}
Comparing different fiber core radii, the maximum coupling efficiency and minimum coupling distance to achieve that efficient coupling are displayed for a range of realistic fiber core sizes in Fig.~\ref{fig:disteffcore}. Our simulations have shown that the maximum coupling efficiency is linearly proportional to the fiber hollow-core area (a quadratic relation with core radius). The minimum coupling distance is also a linear relation of the core size.  For best coupling efficiency, we previously observed that the minimum coupling distance is achieved (with a LG$_{01}$ beam) when the radius of maximum intensity of the diffracting hollow-beam is equal to 1.8 times the diameter of the MOT. Hence, for this fixed funnel radius, the relationship of the distance with the fiber core radius is $d = 1.8 \sigma / tan(\theta)$, where \emph{d} is the coupling distance and \emph{$\theta$} is the diffraction angle of the hollow-beam (inversely proportional to the core radius). The minimum distance is therefore a linear function of the fiber core. With increasing core dimensions, the maximum light force that can be applied is limited by the axial dipolar force (the gravito-optical bottle effect), as reported in Sec.~\ref{sec:cplforce}. It is observed that increasing \emph{$\kappa$} proportionally to $r^3$ will maximize the coupling efficiency. The quadratic increase in the coupling efficiency with core radius (shown in Fig.~\ref{fig:disteffcore}) is a combination of the decreasing axial force by $1/r$ and the linear increase of the gravity potential energy (due to increasing linearly the minimal distance with \emph{r}).

\subsection{\label{sec:loss}Optical scattering in the guiding beam}

We consider the difference in light scattering between atoms guided in a conventional red-detuned Gaussian beam and a blue-detuned hollow-beam. For frequency a detuning large compared to the atomic linewidth and at low saturations, the scattering rate per atom can be expressed:
\begin{equation}
R = \frac{\Gamma }{4}  \frac{ \Omega^{2 }}{\delta^{2}} = \frac{\Gamma^3 }{8\delta^2}\frac{I}{I_s}
\label{eq:lightscattering}
\end{equation}
 
In both guiding arrangements, the cold atoms will be concentrated around the central axis depending on their initial transverse energy. In the blue-detuned guide, the atoms will spend the predominant time in the dark whereas, in the Gaussian beam, they are exposed to the maximum intensity. We have compared the light scattering rates of atoms in repulsive and attractive potentials during the process of coupling and then, guiding inside a HC-PCF.  We have used an identical maximum potential and a detuning of 300\,GHz from resonance in both cases. We have calculated these light scattering rates using two different methods.  

In the first method we explicitly calculate the local intensity, and hence scattering rates, at each time step of our modelled trajectories in the Monte-Carlo simulation, while in the second we have calculated the average light intensity in the volumes explored by atoms of a specific energy and then used this to represent the average scattering rates.  Both techniques predict that only a small fraction of scattering events (less than 8\% of the total) occur during the coupling phase and hence we will ignore this phase in our calculations.  Fig.~\ref{fig:scatters} compares the relationship between light scattering rates and the coupled atom transverse energy, using this first technique, for a Gaussian (black squares) and a LG$_01$ trap beam (green triangles). We observe the expected behavior that a hollow-beam guide has much lower scattering rates than a Gaussian guide except for atoms with a transverse energy close to the guiding potential. 

On Fig.~\ref{fig:scatters} we also display the expected scattering rates from a simple model that averages the beam intensity over the area of confinement of atoms of a particular transverse energy. The radius of that area is determined by a trapping potential that is equal to the transverse energy of the atoms. The calculation of the light scattering rates for the Gaussian (red line) and LG$_01$ beam (blue line) shows excellent agreement with the light scattering rates by the explicit technique. Thus, this investigation shows that blue-detuned hollow-beam guides are superior in the guidance of cold atoms because of the reduced scattering rates while maintaining equivalent guiding potential. It was also observed that for different optical trap depths, the ratio of scattering rates between Gaussian and hollow-beam, for a given transversal energy, is maintained.

 \fig{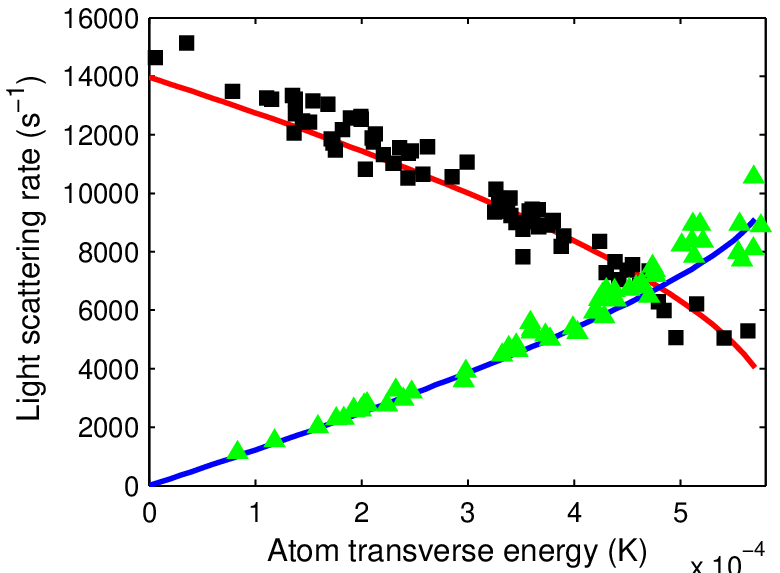}{fig:scatters}{(Color Online) Light scattering rates of cold atoms guided in a HC-PCF by an attractive Gaussian potential (upper red curve and black points) and by a repulsive hollow-core potential (lower blue curve and green points).  We have calculated the expected scattering rates by a statistical method (smooth curves) and by examination of the local intensity through the explicit trajectory of the atoms (points).  The two methods show excellent agreement. Simulated parameters are: 9-mm MOT-fiber distance, 40-$\mu$m core diameter, 0.57\,mK trapping potential depth in the core, and 300\,GHz detuning.}%

\subsection{\label{sec:gausscomp}Model validation and comparison with red-detuned Gaussian beams}

We tested our model against the experimental data of atom guiding efficiency from Mestre \emph{et al.} \cite{Mestre2010}. The Rb atoms were guided using collimated blue-detuned LG modes of order 1 to 12. Because their model didn't take into account light scattering losses, they added an empirical loss factor fitted against their experimental data. We applied the same light scattering loss factor to our simulated results. Table \ref{tab:validation} shows that we obtained excellent agreement with the experimental data. We also applied our model against the cold atom coupling experiment reported recently by Bajcsy \emph{et al.} \cite{Bajcsy2011} (summarized in Sec.~\ref{sec:intro}). We modelled their initial loading configuration as a spatially uniform disk distribution of 440\,$\mu$m diameter, 1\,mm from the fiber, with a temperature distribution equal to that of the MOT. We obtained a coupling efficiency of 1.5\%, with our model compared with their 0.3\% experimentally. Our calculation has not assumed any losses in the MOT loading process and no light scattering losses during the red-detuned Gaussian coupling. 
\begin{table}
\begin{center}
\begin{ruledtabular}
\begin{tabular}{lll}
Beam order & Exp. guiding eff.(\%) & Our model calculation (\%) \\ 
\hline
LG$_0^1$ & 0.8 & 1.4 \\ 
LG$_0^3$ & 7.5 & 8.1 \\ 
LG$_0^5$ & 13.5 & 12.75 \\ 
LG$_0^9$ & 18.0 & 18.3 \\ 
LG$_0^{12}$ & 17.5 & 19.1 \\ 
\end{tabular}
\end{ruledtabular}
\caption{Comparison between experimental data from Mestre \emph{et al.} \cite{Mestre2010} and our model prediction in the guiding efficiency of rubidium atoms in collimated blue-detuned LG beams of various orders.}
\label{tab:validation}
\end{center}
\end{table}

When comparing the coupling efficiency of the blue-detuned hollow-beam with a red-detuned Gaussian beam, the latter has the advantage of a larger confining area (no confinement limits) and the axial dipolar force is attractive, which facilitates coupling. Using the same fiber and identical trap depth potential in the guide, the best coupling efficiency obtained with a red-detuned Gaussian beam has shown to be a factor 4 more efficient (46\%) than the blue-detuned LG$_{01}$ beam (11\%) when using a 40-$\mu$m diameter fiber.

Nonetheless, the blue-detuned hollow-beam is potentially more useful in some circumstances for guidance of cold atoms because of reduced light scattering rates for an identical trapping depth (see Sec.~\ref{sec:loss}).

\section{\label{sec:conclu}Conclusion}

In this paper, we presented a model and simulations aimed at efficient coupling of atoms from a MOT into a hollow-core photonic-crystal fiber using a blue-detuned LG$_{01}$ beam. In contrast with previous approaches, only a single beam is required to achieve both efficient coupling and guiding within the fiber. The key goal is that the atoms are guided in locations with minimal light intensity in order to minimize shifts of the internal states and from decoherence associated with light scattering. 

We showed that, with a blue-detuned hollow-beam funnel, there is a minimum efficient coupling distance, which can be simply calculated from the measured parameters of the MOT and the fiber core radius. Our model identified a new gravito-optical bottle trapping effect that is significant for the coupling dynamics, when using a low-order, blue-detuned hollow-beam funnel. This trapping effect must be carefully managed to optimize coupling efficiency. We identified that optimal atom coupling conditions, for a fixed MOT-fiber separation and fiber core radius, is achieved when using the highest light force possible before the appearance of trapped atoms. This result disputes the naive expectation that strong guiding forces will lead to maximum in-coupling to the fiber.  In contradiction with this expectation, we find that there is an optimal potential depth that will maximize the coupled fraction from the MOT.

When taking into account experimental constraints: a minimum fiber-MOT separation of 9\,mm, hollow-beam power $<$50\,mW; we obtained optimal coupling efficiency of 11\% with a LG$_{01}$ beam when using a 40-$\mu$m core diameter, 25\,mW of power and far-detuning of 3\,THz. Larger core fibers could achieve better coupling efficiency but would require higher light force and experimental constraints would force trading-off power for a reduced detuning, which is not desirable to minimize light scattering losses.


\begin{acknowledgments}
We acknowledge Dr. Boris N. Kuhlmey, Ph.D. (CUDOS, University of Sydney) and Dr. Maryanne C.J. Large, Ph.D. (University of Sydney) for useful discussions on the subject. Jerome Poulin would like to acknowledge the support of the Canadian National Science and Engineering Research  Council (NSERC), Fonds Quebecois pour la Recherche sur la Nature et les Technologies (FQRNT) and the Canadian Institute for Photonics Innovations (CIPI). Raman Kashyap would like to acknowledge the Canada Research Chairs program of NSERC for also supporting this research. We would like to acknowledge the support of the Australian Research Council (ARC). 
\end{acknowledgments}

\bibliography{refs4}

\end{document}